\documentclass[11pt]{article}
\usepackage{amsmath,amsfonts,latexsym,graphicx,amssymb}
\usepackage{amsfonts}
\usepackage{bm}
\newcommand{\In}{{\mathbb{I}_2}}
\newcommand{\IIn}{{\mathbb{I}_{4}}}
\newcommand{\Mn}{M_4(\mathbb{C})}

\newcommand{\ra}{{\, \rightarrow\, }}

\newcommand{\ot}{{\,\otimes\,}}
\newcommand{{\Cd}}{{\mathbb{C}^3}}
\newcommand{\BH}{{\mathcal{B}(\mathcal{H})}}
\newcommand{\BK}{{\mathcal{B}(\mathcal{K})}}
\newcommand{\BHp}{{\mathcal{B}_+(\mathcal{H})}}
\newcommand{\BKp}{{\mathcal{B}_+(\mathcal{K})}}

\def\<{\langle}
\def\>{\rangle}

\newtheorem{thm}{Theorem}[section]
\newtheorem{prop}{Proposition}[section]

\newtheorem{Example}{Example}[section]

\begin{document}

\date{}
%\title{\textbf{On a class of exposed positive decomposable maps }}

\title{\textbf{Exposed positive maps in $\Mn$}}

\author{Dariusz  Chru\'sci\'nski$^1$ and
Gniewomir Sarbicki$^{1,2}$ \\
$^1$ Institute of Physics, Nicolaus Copernicus University,\\
Grudzi\c{a}dzka 5/7, 87--100 Toru\'n, Poland\\
$^2$ Stockholms Universitet, Fysikum, S-10691 Stockholm, Sweden}

\maketitle

\begin{abstract}
It is shown that the family of so called Breuer-Hall maps in $\Mn$ possesses {\em strong spanning property} and hence they are
exposed in the convex cone of positive maps in $\Mn$.
\end{abstract}

\section{Introduction}

Recall that a linear map $\Phi : \BK \ra \BH$ is positive if it maps a cone of positive elements in $\BK$ into  a cone of positive elements in $\BH$, that is, $\Phi(\BHp) \subset \BKp$ \cite{Paulsen}--\cite{Wor}. Throughout this paper we use the standard notation: $\BH$ denotes a $\mathbb{C}^*$-algebra of bounded operators in $\mathcal{H}$ and $\BHp$ denotes a convex cone of positive operators in $\BH$ (recall that $a \in \BHp$ iff $a = bb^*$ for some $b \in \BH$).

Let $\mathcal{P}$ denotes a convex cone of positive maps $\Phi : \mathcal{B}(\mathcal{H}) \longrightarrow \mathcal{B}(\mathcal{K})$ and let $\mathcal{P}^\circ$ denote a dual cone \cite{Eom,Kye-rev,convex}
\begin{equation}\label{}
    \mathcal{P}^\circ = {\rm conv} \{\  P_x \ot P_y\ ;\ \<y|\Phi(P_x)|y\> \geq 0 \ , \ \Phi \in \mathcal{P}\ \}\ ,
\end{equation}
where $P_x = |x\>\<x|$ and $P_y = |y\>\<y|$. It is clear that $\mathcal{P}^{\circ\circ} = \mathcal{P}$, that is, one may consider $\mathcal{P}$ as a dual cone to the convex cone of separable operators in $\mathcal{H}\ot \mathcal{K}$.
Recall that a face of $\mathcal{P}$  is a convex subset $F \subset \mathcal{P}$ such that if the convex combination $\Phi = \lambda \Phi_1  + (1-\lambda)\Phi_2$ of $\Phi_1,\Phi_2 \in \mathcal{P}$ belongs to $F$, then both $\Phi_1,\Phi_2 \in F$.
If a ray $\{ \lambda \Phi\, :\, \lambda > 0\}$  is a face of $\cal P$ then it is called an extreme ray, and we say that $\Phi$ generates an extreme ray. For simplicity we call such $\Phi$ an extremal positive map. A face $F$ is exposed if there exists a supporting hyperplane
$H$ for a convex cone $\mathcal{P}$ such that $F=H \cap \mathcal{P}\,$. The property of `being an exposed face' may be reformulated as follows:
A face $F$ of $\mathcal{P}$ is exposed iff  there exists $a\in \mathcal{B}_+(\mathcal{H})$ and $|h\> \in \mathcal{H}$ such that
$$    F = \{ \ \Phi \in \mathcal{P}\ ; \ \Phi(a)|h\>=0\ \}\ . $$
A positive map $\Phi \in \mathcal{P}$ is exposed if it generates 1-dimensional exposed face.  Let us denote
by ${\rm Ext}(\mathcal{P})$ the set of extremal points and ${\rm Exp}(\mathcal{P})$ the set of exposed points of $\mathcal{P}$. Due to Straszewicz theorem \cite{convex} ${\rm Exp}(\mathcal{P})$  is a dense subset
of ${\rm Ext}(\mathcal{P})$. Thus every extreme map is the limit of some sequence of exposed
maps meaning  that each entangled state may be detected by some exposed positive map.  Hence, a knowledge of exposed maps is crucial for the full characterization of separable/entangled states of bi-partite quantum systems. For recent papers on exposed maps see e.g. \cite{Kye-exposed,GniewkoI,GniewkoIa,Majewski,Marciniak,BH-exp}.

Now, if $F$ is a face of $\mathcal{P}$ then
\begin{equation}\label{dual_face}
    F' = {\rm conv} \{\ a \ot |h\>\<h| \in \mathcal{P}^\circ\ :\ \Phi(a)|h\> = 0 \ , \ \Phi \in F\ \}\ .
\end{equation}
defines a face of $\mathcal{P}^\circ$ (one calls $F'$ a dual face of $F$). Actually, $F'$ is an exposed face. One proves \cite{Eom} the following
\begin{prop}
A face $F$ is exposed iff $F''=F$.
\end{prop}
In this paper we prove the exposedness of a class of positive maps in $\Mn$ using a different tool based on the {\em strong spanning property}.

\section{Exposed maps vs. strong spanning property}

A linear map $\Phi : \BK \ra \BH$ is called {\em irreducible} if  $[\Phi(X),Z] =0$ for all $X \in \BK$ implies $Z = \lambda \mathbb{I}_\mathcal{H}$. In a recent paper \cite{GniewkoI} (see also \cite{GniewkoIa} and \cite{Wor} in connection to unextendible positive maps) we proved the following

\begin{thm}  \label{TH-EX}
Let $\Phi :  \mathcal{B}(\mathcal{K}) \longrightarrow \mathcal{B}(\mathcal{H})$ be a positive, unital, irreducible map, and let
\begin{equation*}\label{}
    N_\Phi = {\rm span} \{ \, a \ot |h\> \in \mathcal{B}_+(\mathcal{K}) \ot \mathcal{H} \ : \ \Phi(a)|h\> = 0\, \}\ .
\end{equation*}
If the subspace $N_\Phi \subset \mathcal{B}(\mathcal{K}) \ot \mathcal{H}$ satisfies
\begin{equation}\label{STRONG}
{\rm dim}\, N_\Phi =  (d_\mathcal{K}^2 -1)d_\mathcal{H}\ ,
\end{equation}
then $\Phi$ is exposed.
\end{thm}
This theorem provides an analog of well known result \cite{Lew} concerning optimality of positive maps (recall that a positive map $\Phi$ is optimal if $\Phi - \Lambda_{CP}$ is no longer positive for any completely positive map $\Lambda_{CP}$)

\begin{thm}  \label{TH-OPT}
Let $\Phi :  \mathcal{B}(\mathcal{K}) \longrightarrow \mathcal{B}(\mathcal{H})$ be a positive map, and let
\begin{equation*}\label{}
    M_\Phi = {\rm span} \{ \, |x\> \ot |h\> \in \mathcal{K} \ot \mathcal{H} \ : \ \Phi(P_x)|h\> = 0\, \}\ .
\end{equation*}
If  $M_\Phi = \mathcal{K} \ot \mathcal{H}$ or equivalently
\begin{equation}\label{WEAK}
{\rm dim}\, M_\Phi =  d_\mathcal{K}d_\mathcal{H}\ ,
\end{equation}
then $\Phi$ is optimal.
\end{thm}
This theorem was used recently to prove the optimality of generalized Choi maps \cite{Filip,Kye-nasza,Gniewko-opt}. In analogy to (\ref{WEAK}) we proposed to call (\ref{STRONG}) {\em strong spanning property}. Hence, as the spanning property is sufficient for optimality the {\em strong spanning property} is sufficient for exposedness.

To illustrate the Theorem \ref{TH-EX} let us consider two simple examples for $\mathcal{K}=\mathcal{H}=\mathbb{C}^2$

\begin{Example}[Transposition map] Let $\tau : M_2(\mathbb{C}) \ra M_2(\mathbb{C})$ denotes the transposition $\tau(X) = X^{\rm t}$ in the standard basis $\{e_1,e_2\}$ in $\mathbb{C}^2$. It is clear that $\tau$ is irreducible and unital. One has $\tau(P_x)|y\>=0$ iff $\< \overline{x}|y\>=0$. The following 6 vectors  $|x_k\> \ot |\overline{x}_k\> \ot |y_k\>$ with
\begin{align*}
& x_1 = e_1\ , \ \ & y_1 =  e_2\ , \\
& x_2 = e_2\ , \ \ & y_2 =  e_1\ , \\
& x_3 = e_1+e_2\ , \ \ & y_3 =  e_1-e_2\ , \\
& x_4 = e_1-e_2\ , \ \ & y_4 =  e_1+e_2\ , \\
& x_5 = e_1+i e_2\ , \ \ & y_5 =  e_1 + ie_2\ , \\
& x_6 = e_1-ie_2\ , \ \ & y_6 =  e_1-e_2\ ,
\end{align*}
are linearly independent in $\mathbb{C}^2 \ot \mathbb{C}^2 \ot \mathbb{C}^2$.  It shows that $\tau$ is an exposed map.
\end{Example}

\begin{Example}[Reduction map] Let $R_2 : M_2(\mathbb{C}) \ra M_2(\mathbb{C})$ denotes the reduction  map
\begin{equation}\label{}
    R_2(X) = \mathbb{I}_2\, {\rm Tr}\, X - X \ .
\end{equation}
Again, one easily shows that $R_2$ is unital and irreducible. Note that
$R_2(P_x)|y\>=0$ iff  $|x\> = \lambda |y\>$ with $\lambda \in \mathbb{C}$. Taking
\begin{equation}\label{}
    x_1 = e_1\ , \ x_2=e_2\ ,\  x_3 = e_1 + e_2\ , \  \ x_4 = e_1 - e_2\ , \ x_5 = e_1 + i e_2\ , \ x_6 = e_1 - i e_2\ ,
\end{equation}
one shows that 6 vectors $|x_k\> \ot |\overline{x}_k\> \ot |x_k\>$ are linearly independent in $\mathbb{C}^2 \ot \mathbb{C}^2 \ot \mathbb{C}^2$. It proves that $R_2$ is an exposed map. Note that $R_n : M_n(\mathbb{C}) \ra M_n(\mathbb{C})$ defined by
\begin{equation}\label{}
    R_n(X) = \mathbb{I}_n\, {\rm Tr}\, X - X \ ,
\end{equation}
is no longer exposed for $n>2$ since it is not extremal.
\end{Example}

\section{Breuer-Hall map in $\Mn$}

Consider now the Robertson map \cite{Robertson} defined by
\begin{equation}\label{}
    \Phi_0(X) = \frac 12 \Big( R_4(X) -  U_0 X^{\rm t} U_0^\dagger \Big) =
    \frac 12 \Big(\mathbb{I}_4\, {\rm Tr}\, X - X - U_0 X^{\rm t} U_0^\dagger \Big) \ ,
\end{equation}
where $U$ is a unitary antisymmetric matrix given by
\begin{equation}\label{}
    U_0 = \left( \begin{array}{c|c} i\sigma_y & \mathbb{O}_2 \\ \hline \mathbb{O}_2 & i\sigma_y \end{array} \right)\ .
\end{equation}
The normalization factor `$1/2$' guaranties that $\Phi_0(\mathbb{I}_4)=\mathbb{I}_4$. It was shown that $\Phi_0$ is an extremal indecomposable map. One has
\begin{equation}\label{}
    \Phi_0(P_x) = \frac 12 \Big(\mathbb{I}_4 - P_x - P_{U\overline{x}} \Big) \ ,
\end{equation}
and $P_x$ and $P_{U\overline{x}}$ are mutually orthogonal projectors  for an arbitrary normalized $x \in \mathbb{C}^4$. Therefore $ \Phi_0(P_x)|y\> =0$ iff $|y\>=|x\>$ or $|y\> = U|\overline{x}\>$.

\begin{prop}
The map $\Phi_0$ is irreducible.
\end{prop}
{\bf Proof}: let us observe that $\Phi_0(X)$ may be rewritten as follows \cite{atomic}
\begin{equation}\label{}
    \Phi_0(X) = \left( \begin{array}{c|c}  \mathbb{I}_2 {\rm Tr}\, X_{22} & - [ X_{12} + R_2(X_{21}) ] \\  \hline
    - [ X_{21} + R_2(X_{12}) ] &  \mathbb{I}_2 {\rm Tr}\, X_{11} \end{array} \right)\ ,
\end{equation}
where $X = \sum_{i,j} e_{ij} \ot X_{ij}\,$ and $e_{ij}$ are matrix units in $M_2(\mathbb{C})$. Suppose now that there exists $Z \in \Mn$ such that $[\Phi_0(X),Z]=0$ for all $X \in \Mn$. Denoting $Y=\Phi_0(X)$ one has
\begin{displaymath}
 %\forall X \in \Im \Phi: \
 \left[ \begin{array}{cc} Y_{11} & Y_{12} \\ \nonumber Y_{21} & Y_{22}\end{array} \right] \cdot
 \left[ \begin{array}{cc} Z_{11} & Z_{12} \\ \nonumber Z_{21} & Z_{22}\end{array} \right] =
 \left[ \begin{array}{cc} Z_{11} & Z_{12} \\ \nonumber Z_{21} & Z_{22}\end{array} \right] \cdot
 \left[ \begin{array}{cc} Y_{11} & Y_{12} \\ \nonumber Y_{21} & Y_{22}\end{array} \right]
\end{displaymath}
implies
\begin{align}
 & Y_{12} Z_{21} = Z_{12} Y_{21} \label{kom1}\ , \\
 & Y_{21} Z_{12} = Z_{21} Y_{12} \label{kom2} \ ,\\
 & Y_{21} Z_{11} + Y_{22} Z_{21} = Z_{21} Y_{11} + Z_{22} Y_{21}\ , \label{kom3} \\
 & Y_{11} Z_{12} + Y_{12} Z_{22} = Z_{11} Y_{12} + Z_{12} Y_{22} \ .\label{kom4}
\end{align}
Note, that if $X_{12}=X_{21}=0$, then necessarily $Z_{12}=Z_{21}=0$ and hence $Z$ is block-diagonal and hence  equations (\ref{kom3}) and (\ref{kom4}) reduce to
\begin{align*}
 & Y_{21} Z_{11} = Z_{22} Y_{21}\ , \\
 & Y_{12} Z_{11} = Z_{22} Y_{12}\ .
\end{align*}
Taking $X_{12}= a\, \In$ and $X_{21} = b\, \In$ with $a,b \in \mathbb{C}$, one gets $Y_{12} = Y_{21} = -(a+b)\In$ and hence $Z_{11}=Z_{22} =: Z_0$. Finally, one obtains the following condition for the diagonal block $Z_0$:
$$ [X_{12} - X_{21},Z_0] = 0\ , $$
%
%\begin{align*}
% & [X_{21} - X_{12} + \Tr X_{12} \In,Z_0] = [X_{21} - X_{12},Z_0] = 0\ , \\
% & [X_{12} - X_{21} + \Tr X_{21} \In,Z_0] =  \ . \\
%\end{align*}
and since $X_{12}$ and $X_{21}$ are arbitrary, it implies $Z_0 = c\, \In$ and hence $Z = c\, \IIn$, which ends the proof of irreducibility of $\Phi_0$.  \hfill  $\Box$

\vspace{.3cm}

\noindent Simple computer algebra enables one to prove the following
\begin{prop} The following 60 vectors $|x_\alpha\> \ot |\overline{x}_\alpha\> \ot |y_\alpha\>$ with $|x_\alpha\>$ belonging to
\begin{eqnarray*}
  x_k &=& e_k \ , \ \ \ (k=1,2,3,4) \\
  x^+_{kl} &=& e_k+e_l\ , (k<l)  \\
  x^-_{kl} &=& e_k-e_l\ , (k<l)\ , (k,l)\neq (1,2)\ , \ (k,l)\neq (3,4) \\
  \widetilde{x}^+_{kl} &=& e_k+ ie_l\ , (k<l)  \\
  \widetilde{x}^-_{kl} &=& e_k- ie_l\ , (k<l)\ , (k,l)\neq (1,2)\ , \ (k,l)\neq (3,4) \\
  x_1' &=& e_1+e_2+e_3 \\
  x_2' &=& ie_1+  e_2+e_3 \\
  x_{3}' &=& e_1+ i e_2+e_3 \\
  x_{4}' &=& e_2+e_3+e_4 \\
  x_{5}' &=& e_2+ie_3+ e_4 \\
  x_{6}' &=& e_2+e_3+ ie_4
\end{eqnarray*}
and $|y_\alpha\> =|x_\alpha\>$ or $|y_\alpha\> = U_0|\overline{x}_\alpha\>$ are linearly independent in $\mathbb{C}^4 \ot \mathbb{C}^4 \ot \mathbb{C}^4$.
\end{prop}
This way we have proved
\begin{thm} The Robertson map $\Phi_0$ is exposed.
\end{thm}

Consider now the family of so called Breuer-Hall map in $\Mn$
\begin{equation}\label{BH}
    \Phi_U(X) = \frac 12 \Big( R_4(X) -  U X^{\rm t} U^\dagger \Big) =
    \frac 12 \Big(\mathbb{I}_4\, {\rm Tr}\, X - X - U X^{\rm t} U^\dagger \Big) \ ,
\end{equation}
where $U$ is an arbitrary unitary antisymmetric matrix. Note, that the above formula defines unital positive map in $M_{n}(\mathbb{C})$ ($n$ even) if $U$ is an antisymmetric unitary from $M_{n}(\mathbb{C})$ and we change the normalization factor $2^{-1} \ra (n-2)^{-1}$.
Clearly, for $U=U_0$ it reproduces the original Robertson map. It was shown \cite{B,H} that $\Phi_U$ is positive and indecomposable. Moreover, it turns out \cite{B} that $\Phi_U$ is nd-optimal, that is, $\Phi_U - \Lambda_D$ is no longer a positive map for an arbitrary decomposable map $\Lambda_D$.

\begin{thm} The map $\Phi_U$ is exposed.
\end{thm}
{\bf Proof}: let us observe that $\Phi_U(\mathbb{I}_4) = \mathbb{I}_4$. 
Note that if the family of vectors
\begin{equation}\label{}
    |x_\alpha\> \ot |\overline{x}_\alpha\> \ot |x_\alpha\> \ , \ \ \
    |x_\alpha\> \ot |\overline{x}_\alpha\> \ot U_0|\overline{x}_\alpha\> \ ,
\end{equation}
spans $\mathbb{C}^4 \ot \mathbb{C}^4\ot \mathbb{C}^4$, then for arbitrary unitary operator $\mathbb{V} : \mathbb{C}^{4 \ot 3} \ra \mathbb{C}^{4 \ot 3}$ the family of vectors
\begin{equation}\label{}
   \mathbb{V}\, ( |x_\alpha\> \ot |\overline{x}_\alpha\> \ot |x_\alpha\>) \ , \ \ \
    \mathbb{V}\, (|x_\alpha\> \ot |\overline{x}_\alpha\> \ot U_0|\overline{x}_\alpha\>) \ ,
\end{equation}
spans $\mathbb{C}^4 \ot \mathbb{C}^4\ot \mathbb{C}^4$ as well. Taking $\mathbb{V} = V \ot \overline{V} \ot V$, where $V : \mathbb{C}^{4} \ra \mathbb{C}^{4}$
is a unitary operator, one finds that the following vectors
\begin{equation}\label{}
    |Vx_\alpha\> \ot |\overline{Vx}_\alpha\> \ot |Vx_\alpha\> \ , \ \ \
    |Vx_\alpha\> \ot |\overline{Vx}_\alpha\> \ot U|\overline{Vx}_\alpha\> \ ,
\end{equation}
with
\begin{equation}\label{UU0}
    U = V U_0 V^{\rm t}\ ,
\end{equation}
span $\mathbb{C}^4 \ot \mathbb{C}^4 \ot \mathbb{C}^4$. Note, that $V U_0 V^{\rm t}$ defines antisymmetric unitary for an arbitrary unitary $V$. Moreover, any
antisymmetric unitary $U$ may be written via (\ref{UU0}) for an appropriate unitary $V$, that is, $U_0$ is a canonical form for an antisymmetric unitary matrix in $\Mn$. Indeed, note that for any antisymmetric  unitary $U$ if $\lambda$ is an eigenvalue so is $-\lambda$ and $\lambda = e^{i\alpha}$. It is, therefore, clear that
\begin{equation}\label{}
    U = R\, {\rm diag} \{ e^{i \alpha_1} i\sigma_y, e^{i \alpha_2} i\sigma_y \}\, R^{\rm t}\ ,
\end{equation}
where $R \in M_4(\mathbb{R})$ is an orthogonal matrix. It proves that $\Phi_U$ possesses the strong spanning property for an arbitrary antisymmetric unitary $U$. It remains to show that $\Phi_U$ is irreducible. We have already proved that
$[\Phi_{0}(X),Z]=0$ for all $X \in \Mn$ implies $Z \sim \mathbb{I}_4$. Note, that
$$ 0 = [\Phi_{0}(X),Z] = V^\dagger [V\Phi_{0}(X)V^\dagger,VZV^\dagger] V = V^\dagger[ \Phi_U(X'),Z']V\ , $$
where $ U = VU_0V^{\rm t}$, $X'= V^\dagger X V$ and $Z'=VZV^\dagger$. Now, $[ \Phi_V(X'),Z'] =0$ for all $X'\in \Mn$ implies $Z' \sim \mathbb{I}_4$ which proves irreducibility of $\Phi_U$. 

Hence, $\Phi_U$ possesses the {\em strong spanning property} for an arbitrary $U$ and being  irreducible  it is an exposed map. \hfill $\Box$

\section{Conclusions}

We have shown that so called Breuer-Hall positive maps \cite{B,H} in $\Mn$  satisfies the {\em strong spanning property} and hence they are exposed in the convex cone of positive maps in $M_{4}(\mathbb{C})$. Therefore, this class defines the most efficient tool for detecting quantum entanglement (any entangled state may be detected by some exposed map (entanglement witness). In a recent paper \cite{BH-exp} it was shown that Breuer-Hall maps in $M_{n}(\mathbb{C})$ (and $n$ even) are exposed for an arbitrary antisymmetric unitary $U$. Interestingly, numerical analysis shows that for $n>4$ these maps do not possesses the {\em strong spanning property}. Hence, for $n>4$ they provide an analog of the Choi map which is known to be optimal (even extremal) but does not satisfy spanning property (\ref{WEAK}). Actually, numerical analysis shows that
$$ D_n:= {\rm dim}\, N_\Phi = \frac 16 n (n+1)(5n-2) \ . $$
Note that
$$ D_n \leq n(n^2-1)\ , $$
and the equality holds only for $n=4$.

\section*{Acknowledgments}

G.S. was partially supported  by research fellowship within project {\em Enhancing Educational Potential of Nicolaus Copernicus University in the Disciplines of Mathematical and Natural Sciences}   (project no. POKL.04.01.01-00-081/10.)

\end{document}